\documentclass[twocolumn]{autart} 
\pdfoutput=1

\usepackage[usenames, dvipsnames]{color}
\usepackage{soul}
\usepackage{xcolor}
\usepackage{amssymb}
\usepackage{mathtools}
\usepackage{enumerate}
\usepackage{amsmath}
\usepackage{array}
\usepackage{algorithm}
\usepackage{subcaption}
\usepackage{balance}
\usepackage[noend]{algpseudocode}
\usepackage[normalem]{ulem}
\useunder{\uline}{\ul}{}
\usepackage{pdfpages}
\makeatletter
\newcommand*{\rom}[1]{\expandafter\@slowromancap\romannumeral #1@}
\makeatother
\makeatletter
\algrenewcommand\algorithmiccomment[2][\normalsize]{{#1\hfill\(\triangleright\) #2}}
\makeatother
\newtheorem{theorem}{Theorem}

\newtheorem{definition}{Definition}

\newif\ifComments
\Commentstrue

\usepackage{graphicx}
\graphicspath{{Images/}}
\DeclareGraphicsExtensions{.pdf,.emf,.png}
\usepackage{natbib}        

\DeclareRobustCommand{\hleme}[1]{{\sethlcolor{OliveGreen}\hl{#1}}}
\DeclareRobustCommand{\hlbred}[1]{{\sethlcolor{BrickRed}\hl{#1}}}
\DeclareRobustCommand{\hlmblue}[1]{{\sethlcolor{MidnightBlue}\hl{#1}}}

\begin{document}

\begin{frontmatter}

\title{A Tree Adjoining Grammar Representation for Models Of Stochastic Dynamical Systems\thanksref{footnoteinfo}\thanksref{footnoteinfo2}} 

\thanks[footnoteinfo]{This research is supported by the Dutch Organization for Scientific Research (NWO, domain TTW, grant: 13852) which is partly funded by the Ministry of Economic Affairs of The Netherlands.}
\thanks[footnoteinfo2]{Corresponding author D.~Khandelwal.}

\author[a]{Dhruv Khandelwal}\ead{D.Khandelwal@tue.nl},    
\author[a]{Maarten Schoukens}\ead{M.Schoukens@tue.nl},               
\author[a]{Roland T\'oth}\ead{R.Toth@tue.nl}  

\address[a]{Department of Electrical Engineering, Eindhoven University of Technology, Eindhoven, The Netherlands}  


\begin{abstract}                          
Model structure and complexity selection remains a challenging problem in system identification, especially for parametric non-linear models. Many Evolutionary Algorithm (EA) based methods have been proposed in the literature for estimating model structure and complexity. In most cases, the proposed methods are devised for estimating structure and complexity within a specified model class and hence these methods do not extend to other model structures without significant changes. In this paper, we propose a Tree Adjoining Grammar (TAG) for stochastic parametric models. TAGs can be used to generate models in an EA framework while imposing desirable structural constraints and incorporating prior knowledge. In this paper, we propose a TAG that can systematically generate models ranging from FIRs to polynomial NARMAX models. Furthermore, we demonstrate that TAGs can be easily extended to more general model classes, such as the non-linear Box-Jenkins model class, enabling the realization of flexible and automatic model structure and complexity selection via EA.
\end{abstract}

\end{frontmatter}

\section{Introduction}

\vspace*{-0.2cm}
In recent years, there has been a resurgence in the use of Evolutionary Algorithms (EAs) for data-driven modelling of dynamical systems. Undoubtedly, one of the main driving forces for this is the steady growth of computation power. EAs are being increasingly used in a multitude of engineering domains and life science \citep{eiben2003introduction,arias2012multiobjective}. Across several domains, EAs have generated results that are competitive and, sometimes, even surprising \citep{eiben2003introduction}. Another factor contributing to the growing popularity of EAs is that these algorithms can be used to generate solutions for complex problems for which no systematic solution approach exists in general. In parametric system identification, the estimation of model structure and model complexity is one such problem.

Model structure selection is a classical problem in system identification. Over the years, a variety of methods for system identification have been developed. Each of these methods adopt different approaches to solve the problem of model structure selection. While methods like Prediction Error Minimization (PEM) treat model structure selection as a user's choice \citep{Ljung1999}, other methods (for example, \cite{pillonetto2011prediction}, \cite{laurain2020sparse}) rely on a flexible model structure, and attempt to estimate or control the complexity of the model-to-be-estimated via regularization. Furthermore, the appropriate model complexity is often chosen by ranking models based on an information metric, such as AIC, BIC, or based on a user-defined complexity measure \citep{rojas2014sparse}. In cases where the number of candidate models grows combinatorially with respect to the length (or the complexity) of the model, a ranking-based complexity selection strategy becomes intractable, restricting model structure selection to regularization or shrinkage based methods. 

As a consequence of the aforementioned challenges, heuristics-based methods such as EAs have been used to estimate model structure and complexity, with a fair amount of success. However, the application of EAs have been, to some extent, superficial. The premise of the biologically-inspired heuristics used in EAs is that the solutions of a given problem can be constructed from fundamental building blocks, and these fundamental components can be interchanged between different solutions. In the system identification literature, the proposed EA-based approaches to model structure and complexity selection can be categorized as follows:
\begin{enumerate}
	\item[i)] approaches that choose a fixed model structure and use EAs to determine the appropriate model complexity (or model terms), and
	\item[ii)] approaches that use EAs to explore model structure and model complexity.
\end{enumerate}
In the first category of EA-based approaches, the basic building blocks of an EA are chosen such that only models with a specific model structure can be generated. Hence, these approaches cannot be typically extended to other model structures without significant modifications. This approach can be found in \cite{fonseca1996non,rodriguez2004identifying,rodriguez2000use}, where the authors use EAs to perform term selection within a chosen model structure. This approach is also used in \cite{kristinsson1992system}, where the authors use GAs to estimate pole-zero locations for ARMAX models.

In the second category of EA-based approaches, more generic set of building blocks are used in the EA, allowing the generation of models with arbitrary model structures. In this case, EAs are used to determine not just the appropriate complexity of the model, but also the approprite model structure (e.g., in terms of the non-linear functions to be included in the model). However, unrestrained generation of arbitrary model structures using EA may result in models that are not well-posed, e.g., models with discontinuities, non-causality, or finite escape-time. Typically, these problems are avoided by using arbitrary ad-hoc solutions, e.g, setting all discontinuities to 0. Another common drawback of EA-based approaches that fall in the second category is that prior knowledge of the dynamical system cannot be incorporated systematically in the identification procedure. In \cite{madar2005genetic}, the authors use GP to identify NARMAX models that may contain arbitrary non-linearities. While the authors are interested in models that are linear-in-the-parameters, GP may return models that do not belong to that class. Consequently, the authors use an ad-hoc solution to ensure that the candidate model structures generated by GP are linearly parameterized. A similar approach was used in \cite{quade2016prediction} with a larger set of mathematical operations. Again, the proposed approach does not allow for systematic inclusion of model structure constraints or prior knowledge of the system. A slightly different approach is used in \cite{gray1998nonlinear}, where the authors use GP to construct linear or non-linear models from basic elements like SIMULINK blocks and static non-linearities. Again, the combination of various SIMULINK blocks cannot be systematically structured to avoid ill-posed models.

In this paper, we propose a generative grammar based representation of stochastic parametric dynamical systems. The proposed representation allows for the generation of complex, yet well-posed dynamical models by combining a set of fundamental building blocks in well-specified ways. The resulting generative declaration of models defines a notion of \emph{model set} that is more generalized than that conventionally used, for example, in \cite{Ljung1999}. The generative grammar used in this work is called Tree Adjoining Grammar (TAG) \citep{joshi1997tree}. The use of TAG in an EA-based approach makes it possible to develop a system identification framework where EAs are used to automatically determine the structure and complexity of a model from a generic, well-posed class of dynamical models, while systematically incorporating model structure constraints and prior knowledge. A preliminary concept of the proposed framework (without proofs) was presented in \cite{khandelwal2019grammar}. The proposed approach for grammar-based identification was found to produce results that were comparable to state-of-the-art non-linear system identification approaches, while using no specialized knowledge of the benchmark system being identified.

The main contributions of this paper are the following. We present a detailed discussion on the discrete-time input-output representation of dynamical systems using TAG, and introduce a new notion of a model set defined by the generative capacity of a TAG. Subsequently, we develop a TAG for the polynomial NARMAX model class. We prove that any model structure generated by the proposed TAG belongs to the class of polynomial NARMAX models, and conversely, any polynomial NARMAX model can be represented using the proposed TAG (for which an algorithm is also proposed). We demonstrate that the model set corresponding to the proposed TAG includes, as special cases, other commonly used model structures such as FIR, ARX and Truncated Volterra series models. We also demonstrate that the proposed representation can be easily extended to other model structures (namely polynomial Non-linear Box-Jenkins, or NBJ). Note that, while the TAG-based model set notion developed in this contribution is motivated by its applicability in an EA-based identification methodology, the identification approach itself is \emph{not} in the scope of the present contribution. A preliminary version of such an identification methodology can be found in \cite{khandelwal2019data} and \cite{khandelwal2019grammar}.

The contributions in this paper differ from \cite{khandelwal2019grammar} in the following respects:
\begin{itemize}
	\item we formulate a TAG for a larger class of dynamical systems (the polynomial NARMAX class), and prove their equivalence,
	\item we provide an algorithm to compute an equivalent TAG representation of a given polynomial NARMAX model,
	\item we illustrate, via examples, the restriction (and generalization) of the proposed TAG in order to generate models with more specific (or generic) structures.
\end{itemize}

The remainder of the paper is structured as follows. The concept of TAG is introduced, both informally and formally, in Sec. \ref{sec:TAG}. In Sec. \ref{sec:dynamical_TAG} we introduce the notion of model set as defined by a given TAG, and propose a TAG that generates the class of polynomial NARMAX models. Several examples are used to illustrate the concept in Sec. \ref{sec:illustrations}, followed by concluding statements in Sec. \ref{sec:conclusions}.

\vspace*{-0.15cm}
\section{Tree Adjoining Grammar}
\label{sec:TAG}
\vspace*{-0.1cm}

To set the stage for the development of TAG for stochastic non-linear systems, first we introduce the basic concepts of TAG. Since TAG was initially developed from linguistic considerations, a linguistic example will be used to illustrate the methodology. This will be followed by formal definitions. To make the example illustrative, we first specify an example string, and then infer a TAG that would generate the given string. Conversely, for the formal definitions, we will begin with the basic components of a TAG and lead up to the definition of TAG and operations that can be performed on TAGs.
	
	\vspace*{-0.15cm}
	\subsection{An informal description}
	\vspace*{-0.15cm}
	Informally, a \textit{formal grammar} can be described as a set of rules for generating strings. The resulting set of strings is called the \textit{language} generated by the grammar. In contrast, TAG describes a set of rules for generating trees. The resulting set of trees is called the \textit{tree language} of the TAG. The \textit{yield} of all the trees in the tree set subsequently determines the corresponding language.
	
\begin{figure} [!tb]
			\centering
			\includegraphics[width = 1.7in]{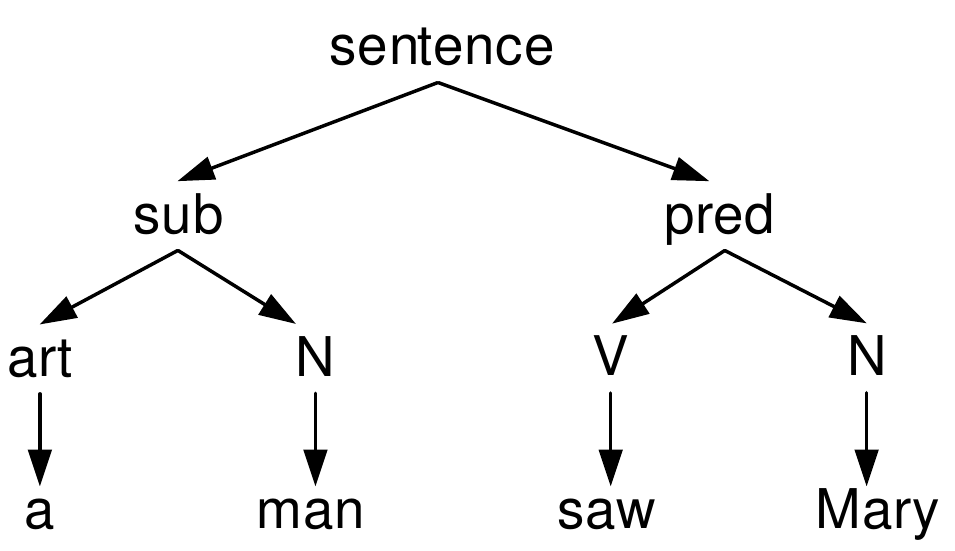}
			\caption{A derived tree with the yield "A man saw Mary". The tree depicts the grammatical constructs that are evident in the structure of the sentence - a subject (sub) and a predicate (pred), an article (art), a verb (V) and nouns (N).}
			\label{fig:sentenceDerivedTree}
	\end{figure}

	The following example has been derived from \cite{joshi1997tree}. Consider the sentence ``A man saw Mary''. Simple grammatical constructs can be used to decompose the given sentence into its basic components. For example, the sentence consists of articles (``A''), nouns (``man'', ``Mary'') and verbs (``saw''). Other underlining structures, such as subjects and predicates, can also be observed in the sentence. The sentence, together with the underlying grammatical structure can be represented in a single tree structure as shown in Fig. \ref{fig:sentenceDerivedTree}. The tree depicted in Fig. \ref{fig:sentenceDerivedTree} is called a \textit{derived tree}. The \textit{yield} of a derived tree are the labels associated with the leaves of the tree. Hence, the yield of the derived tree in Fig. \ref{fig:sentenceDerivedTree} is ``A man saw Mary''.
	
\begin{figure} [!tb]
		\centering
		\begin{subfigure}[t]{0.6\linewidth}
			\includegraphics[width = 2.1in]{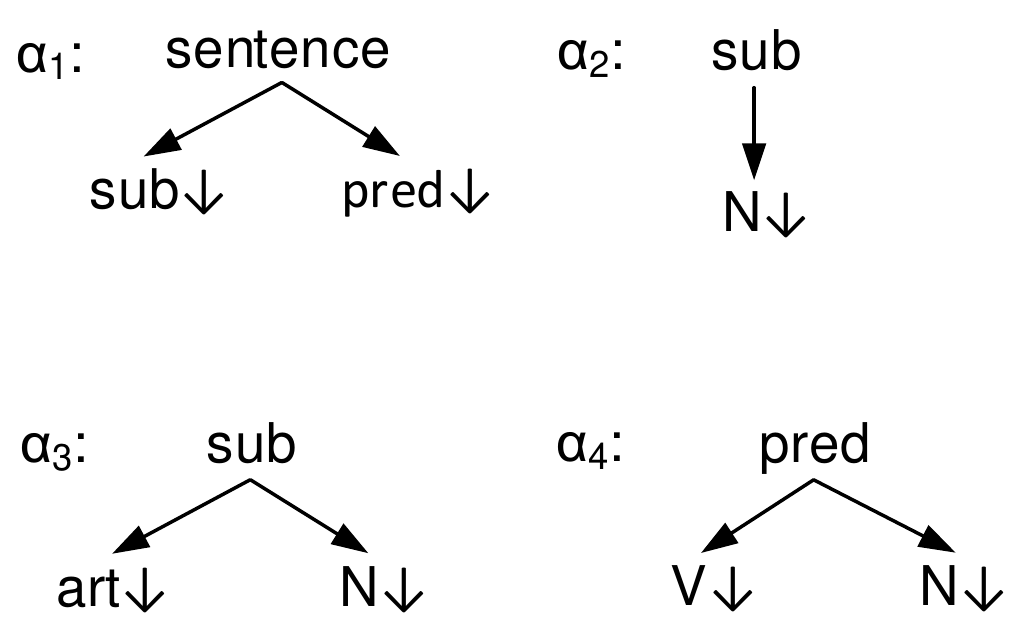}
			\caption{Set of \textit{initial trees} ($I$).}
			\label{fig:sentenceI}
		\end{subfigure}
		~%
		\begin{subfigure}[t]{0.37\linewidth}
			\includegraphics[width = 1.3in]{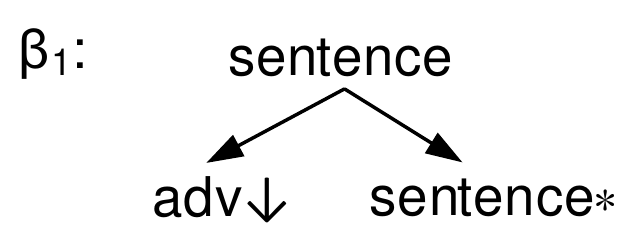}
			\caption{Set of \textit{auxiliary trees} ($A$).}
			\label{fig:sentenceA}
		\end{subfigure}
		\caption{The sets $I$ and $A$ serve as building blocks of the tree set of a TAG.}
		\label{fig:sentenceIandA}
	\end{figure}

	The given derived tree can be obtained by combining basic building blocks that are constituents of the TAG. Fig. \ref{fig:sentenceIandA} depicts the set of initial trees $I$ and auxiliary trees $A$, collectively known as \emph{elementary trees}, that can be combined in specific ways to produce the derived tree in Fig. \ref{fig:sentenceDerivedTree}. The set of initial trees $I$ can be informally described as a set of non-recursive replacement rules that can be used to generate a set of trees. The set of auxiliary trees can be described as a set of recursive replacement rules. Consequently, each auxiliary tree has a terminal node with the same label as that of its root node.
	
	The downward arrow symbol $\downarrow$ and the star symbol $\star$ in Fig. \ref{fig:sentenceIandA} represent nodes in a tree that are available for a \textit{substitution} and \textit{adjunction} operation respectively. A substitution operation can be used to substitute an initial tree into, for instance, another initial tree, if and only if the latter has a terminal node (leaf) with a label that matches the label of the root node of the prior. On the other hand, adjunction can be loosely described as the operation of inserting an auxiliary tree into a syntactic tree. Adjunction of an auxiliary tree can take place on a non-terminal node of a syntactic tree if and only if the node has a label that matches the label of the root node of the auxiliary tree to be adjoined.
	
\begin{figure} [!tb]
		\centering
		\begin{subfigure}[t]{0.45\linewidth}
			\centering
			\includegraphics[width = 0.7 in]{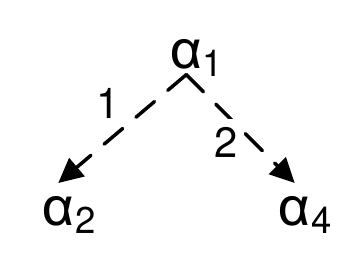}
			\caption{Derivation tree for "a man saw Mary".}
			\label{fig:sentenceDerivation1}
		\end{subfigure}
		~%
		\begin{subfigure}[t]{0.45\linewidth}
			\centering
			\includegraphics[width = 1.1 in]{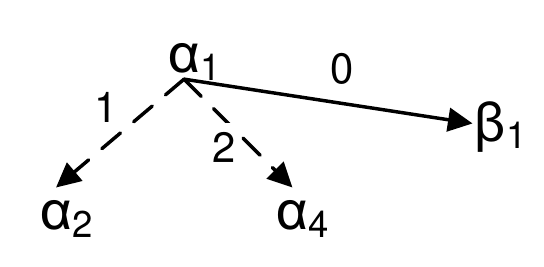}
			\caption{Derivation tree for "yesterday a man saw Mary".}
			\label{fig:sentenceDerivation2}
		\end{subfigure}
		\caption{Derivation tree representation - dashed lines represent substitutions, solid lines represent adjunction, and labels on the edges represent the Gorn addresses (a method to assign a label to a node in a tree structure, see \cite{gorn1965explicit}) of the nodes participating in substitution or adjunction.}
		\label{fig:sentenceDerivations}
	\end{figure}

	Consider the following sequence of operations. The initial tree $\alpha_3$ can be substituted in $\alpha_1$ at the location of the ``sub'' node.  Let's denote the resulting tree as $\gamma_1$. The tree $\gamma_1$ is an example of a \textit{syntactic tree}, a tree obtained by applying an arbitrary number of substitution and adjunction operations to a given initial tree. Again, the initial tree $\alpha_4$ can be substituted to the syntactic tree $\gamma_1$ at the location of the ``pred'' node. Let the result be denoted as $\gamma_2$. Note that $\gamma_2$ has the same structure as the example in Fig. \ref{fig:sentenceDerivedTree}, upto the last level of the derived tree, where specific articles, nouns and verbs are substituted in the tree to obtain the yield ``a man saw Mary''. Substitution can be performed on a initial tree or syntactic tree as long as there exist nodes available for substitution, marked by $\downarrow$. A \textit{derived tree} is a syntactic tree in which none of the terminal nodes (leaves) are available for substitution. The initial and auxiliary trees provide an alternative representation, the \textit{derivation tree}, as shown in Fig. \ref{fig:sentenceDerivation1}. Based on the TAG in Fig. \ref{fig:sentenceIandA}, more complex sentences can also be generated. For example, the auxiliary tree $\beta_1$ can be adjoined to the root node of $\gamma_2$ since both root nodes have the label ``sentence''. This operation effectively adds an adverb before the sentence, yielding the sentence ``yesterday a man saw Mary''. The resulting derivation tree is depicted in Fig. \ref{fig:sentenceDerivation2}.
	
	The set of all derived trees that can be obtained, by starting from a given start symbol, say ``sentence'', and applying an arbitrary number of adjunctions and/or substitutions using elementary trees is called the \textit{tree language} of the corresponding TAG. The string yield of all trees in the tree set is called the \textit{string language} of the corresponding TAG.
	
	We can now introduce the formal definitions of the concepts that were informally described in this example.
	
	
	\subsection{The formal definitions}
	\vspace*{-0.15cm}
	The formal definitions of TAG and related concepts can be found in \cite{joshi1997tree} and \cite{kallmeyer2009declarative}. These definitions are reproduced here for completeness.
	
	\begin{definition}
		A \textit{finite tree} is a directed graph, denoted by $\gamma =<V,E,r>$, where, $V$ is the set of vertices, $E$ is the set of edges, and $r \in V$ is the root node, such that
		\begin{itemize}
		\item[-] $\gamma$ contains no cycles,
		\item[-] $r \in V$ has in-degree (number of incoming edges) 0,
		\item[-] All $v \in V \setminus \{r\}$ have in-degree 1,
		\item[-] Every $v \in V$ is accessible from $r$,
		\item[-] A vertex with out-degree (i.e., number of outgoing edges) 0 is a leaf. 
		\end{itemize}
	\end{definition}
	
	\begin{definition}
		A \textit{labeling} of a graph $\gamma = <V,E>$ over a \textit{signature} $<A_1,A_2>$ is a pair of functions $l:V\rightarrow A_1$ and $g:E \rightarrow A_2$, with $A_1,A_2$ being a set of disjoint alphabets.
	\end{definition}
	
	For the next definitions, assume $N$ and $T$ to be disjoint sets of non-terminals and terminals, respectively.
	
	\begin{definition}
		A \textit{syntactic tree} is an ordered, labelled tree ${<V,E,r>}$ such that the label $l(v) \in N$ for each vertex $v$ with out-degree at least 1 and $l(v) \in (N \cup T \cup \epsilon)$ for each leaf $v$.
	\end{definition}
	
	\begin{definition}
		An \textit{auxiliary tree} is a syntactic tree ${<V,E,r>}$ such that there is a unique leaf $f$, marked as \textit{foot node}, with $l(f)=l(r)$. An auxiliary tree is denoted as ${<V,E,r,f>}$.
	\end{definition}
	
	\begin{definition}
		An \textit{initial tree} is a non-auxiliary syntactic tree.
	\end{definition}
	
	With the basic concepts defined, we can now define TAG, and the related operations.
	
	\begin{definition}
	 A \textit{Tree Adjoining Grammar} is a tuple ${G = <N,T,S,I,A>}$, where
	 
	 \begin{itemize}
	 	\item[-] $N,T$ are disjoint alphabets of non-terminals and terminals,
	 	\item[-] $S\in N$ is a start symbol,
	 	\item[-] $I$ is a finite set of initial trees and $A$ is a finite set of auxiliary trees.
	 \end{itemize}
	\end{definition}
	
	The set of trees $I \cup A$ is called \emph{elementary trees}.
	
	\begin{figure}
		\vspace*{0.2cm}
		\centering
		\begin{subfigure}[t]{0.9\linewidth}
			\includegraphics[scale = 0.37]{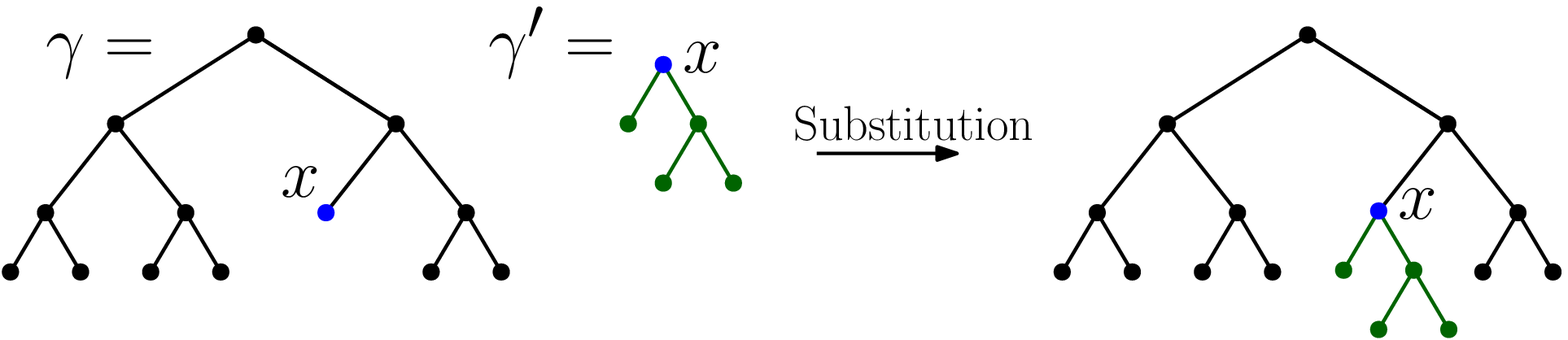}
			\caption{TAG substitution operation.}
			\label{fig:substitution}
		\end{subfigure}
		\\
		\begin{subfigure}[t]{0.9\linewidth}
			\includegraphics[scale = 0.37]{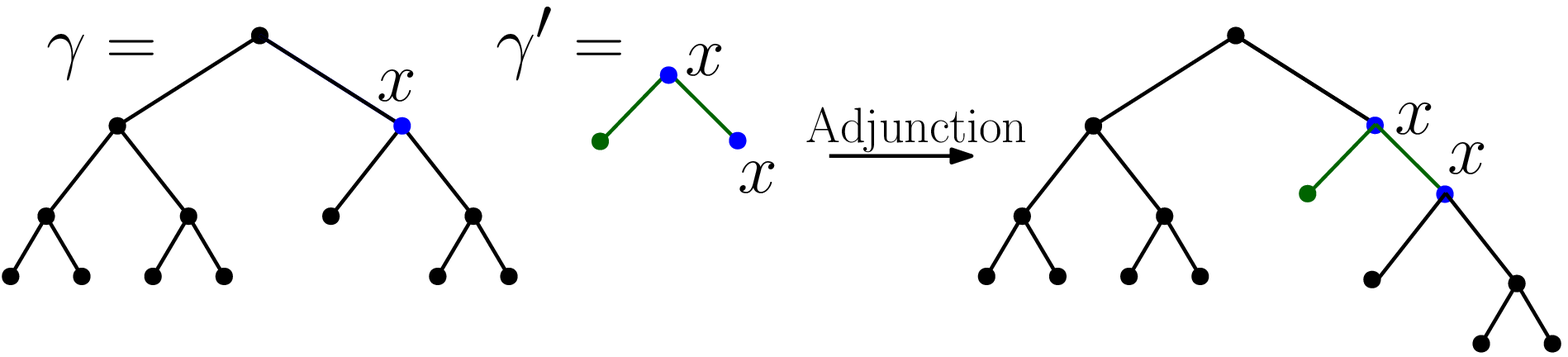}
			\caption{TAG adjunction operation.}
			\label{fig:adjunction}
		\end{subfigure}
		\caption{Illustration of the TAG operations \citep{khandelwal2019grammar}.}
		\label{fig:operations}
\end{figure}
	\begin{definition}[Substitution]
		Let $\gamma = {<V,E,r>}$ be a syntactic tree and $\gamma' = {<V',E',r'>}$ be an initial tree and $v \in V$. The result of substituting $\gamma'$ into $\gamma$ at node $v$, denoted as $\gamma[v,\gamma']$, is defined as follows
		\begin{itemize}
			\item[-] If $v$ is not a leaf or $v$ is a foot node or $l(v) \ne l(r')$, then $\gamma[v,\gamma']$ is not defined,
			\item[-] otherwise, $\gamma[v,\gamma'] = {<V'',E'',r>}$ with 
			\begin{equation}
			V'' = {V \cup V' \setminus \{v\}},
			\end{equation}
			and
			\begin{multline}
			E'' = \left(E \setminus \{{<v_1,v_2>} \mid v_2=v \text{ and } v_1 \in V\} \right) \ \cup  \\
			 E' \ \cup \ \{{<v_1,r'>} \mid {v_1,v} \in E \}.
			\end{multline}
		\end{itemize}
	\end{definition}
	The substitution operation is illustrated in Fig. \ref{fig:substitution}.
	
	\begin{definition}[Adjunction]
		Let $\gamma = {<V,E,r>}$ be a syntactic tree and $\gamma' = {<V',E',r',f>}$ be an auxiliary tree and $v \in V$ with out-degree at least 1. The result of adjoining $\gamma'$ into $\gamma$ at node $v$, denoted as $\gamma[v,\gamma']$, is defined as follows
		\begin{itemize}
			\item[-] if $l(v) \ne l(r')$ then $\gamma [ v,\gamma']$ is undefined,
			\item[-] else $\gamma[v,\gamma'] = {<V'',E'',r''>}$ with
			\begin{equation}
				V'' = V \cup V'\setminus {v},
			\end{equation}
			and
			\begin{multline}
				E''= \left( E \setminus \left\{ {<v_1,v_2>} \mid v_1 = v \text{ or } v_2 = v\right\} \right) \ \cup \\
				E' \ \cup \ \left\{ {<v,r'>} \mid {<v_1,v>} \in E \right\} \ \cup \\
				\left\{ {<f,v_2>} \mid {<v,v_2>} \in E \right\}.
			\end{multline}
		\end{itemize}
	\end{definition}
	The adjunction operation is illustrated in Fig. \ref{fig:adjunction}.
	
	Recall that a tree obtained by performing an arbitrary number of valid substitution and adjunction operations to an initial tree $\gamma = \left< V, E, r \right>$ with $l(r) = S$ is called a \textit{derived tree} (for example, as in Fig. \ref{fig:sentenceDerivedTree}). Also recall that the substitution and adjunction operations performed can be represented in a tree representation called \textit{derivation tree} (for example, as in Fig. \ref{fig:sentenceDerivations}). A derived tree is said to be \textit{saturated} if all leaves of the derived tree belong to the set $T$ and cannot be further substituted. The corresponding derivation tree is also said to be \textit{saturated}.
	
	\begin{definition}[Tree language and string language]
		Let $G = \left< N,T,S,I,A \right>$ be a TAG. The \textit{tree language} $L_\mathrm{T}(G)$ of grammar $G$ is defined as the set of all saturated derived trees in $G$ with root $S$.
		
		The \textit{string language} $L(G)$ of $G$ is the set of yields of the trees in $L_\mathrm{T}(G)$.
	\end{definition}
	
\vspace*{-0.15cm}
\section{TAG Description of Dynamical Systems}
\label{sec:dynamical_TAG}
\vspace*{-0.1cm}
In this Section, we define a notion of \textit{model set} based on TAG and propose a TAG for a generic class of dynamical models - the polynomial NARMAX class.

\vspace*{-0.15cm}
\subsection{Model set}
\vspace*{-0.15cm}
Consider the following discrete-time input-output representation of a non-linear dynamical model
\begin{multline}
	y_k = f(u_k,\dots, u_{k-n_u},y_{k-1},\dots, y_{k-n_y}, \xi_{k-1},\dots, \\ \xi_{k-n_\xi}) + \xi_k \label{eq:IO}
\end{multline}
where $u_k,y_k \in \mathbb{R}$ are the input and output signals at time-instant $k$, $\xi_k \sim \mathcal{N}(0, \sigma_\xi^2)$ is a noise signal independent of input $u$, constants $n_u, n_y$ and $n_\xi$ are the corresponding maximum time-lags and the non-linear function $f(\cdot)$ belongs to an arbitrary set of functions $\mathcal{M}$. In PEM, the set of functions $\mathcal{M}$, also known as the \textit{model set}, along with a specified choice for $n_u, n_y$ and $n_\xi$, is determined by a user based on expert knowledge, prior information and informative experiments. It will be demonstrated in Sec. \ref{sec:NARMAX_TAG} that TAG can be used to generate trees that yield non-linear functions $f(\cdot)$ with desirable structural properties and varying choices of arguments (time lags of the involved $u,y$ and $\xi$ signals). This capability of TAG leads to a more generalized notion of model set $\mathcal{M}$. In order to formalize this concept, we introduce a function $\Pi_f(u,y,\xi,k)$ that maps from function $f$ to the right-hand-side expression in \eqref{eq:IO} (in string form). We can now define a new notion of model set, based on TAG, defined as follows.
	\begin{definition}
		For a given TAG $G$, the corresponding \textit{model set} $\mathcal{M}(G)$ is defined as the set of models in the form of \eqref{eq:IO} such that $\Pi_f(u,y,\xi,k)\in L_\mathrm{T}(G)$.
		\label{C4:def:modelSet}
	\end{definition}
Note that this is a more generalized notion of model set as compared to that used in PEM. In PEM, a model set is typically determined by choosing a fixed model structure along with a suitable parameterization (i.e. model complexity). On the other hand, in this work, the choice of initial and auxiliary trees of a TAG automatically determines the model set. The advantage of such a declaration of a model set is that, when no prior information is available, the model set can be chosen to span a number of commonly used model classes without a prior specification of the model complexity. On the other hand, when prior information on the structure or complexity of the model is available, the grammar can be suitably refined to restrict the model set. In the subsequent sections, we propose a TAG for a generic model class, and demonstrate that the resulting model set spans a number of model structures commonly used in PEM.

\vspace*{-0.15cm}
\subsection{The polynomial NARMAX model class}
\label{sec:NARMAX_TAG}
\vspace*{-0.15cm}

The NARMAX model class is a flexible class on non-linear input-output dynamical models, see \cite{leontaritis1985input}. The polynomial NARMAX model class is the set of all NARMAX models where the non-linear relationships are of the polynomial kind. Polynomial NARMAX is a convenient model representation since any continuous function on a closed space can be approximated arbitrary well using polynomial functions (based on Weierstrass' theorem, see \cite{stone1948generalized}). Furthermore, the family of polynomial NARMAX models includes, as special cases, other commonly used model classes such as FIR and ARMAX. It will be shown that these models can be generated by suitably restricting the TAG presented here.

A discrete-time SISO polynomial NARMAX model can be represented as (see \cite{billings2013nonlinear})
\vspace*{-0.10cm}
\begin{multline}
	y_k = \theta_0 + \sum_{i_1 = 1}^{n} \theta_{i_1} x_{i_1,k} + \sum_{i_1 = 1}^{n} \sum_{i_2 = i_1}^n \theta_{i_1 i_2} x_{i_1,k} x_{i_2,k} + \dots \\
	\sum_{i_1 = 1}^n \dots \sum_{i_l = i_{l-1}}^n \theta_{i_1 i_2 \dots i_l} x_{i_1,k} x_{i_2,k} \dots x_{i_l,k} + \xi_k, \label{eq:NARMAX1}
\end{multline}
where $l$ is the order of the polynomial non-linearity, 
$\theta_{i_1 i_2 \dots i_m}$ are the model parameters, and $x_{k} = (x_{1,k} \cdots $ $ x_{n_y + n_u + n_\xi,k})^\top $ is a vector consisting of the past input, output and noise values building up the regressors

\vspace*{-0.5cm}
{\small
\begin{equation}
	x_{m,k} =  \begin{cases}
	y_{k-m} & 1 \leq m \leq n_y\\
	u_{k - (m - n_y - 1)} & n_y + 1 \leq m \leq n_y + n_u + 1 \\
	\xi_{k - (m - n_y - n_u - 1)} & \begin{split}
	n_y + n_u + 2 \leq m \leq n_y + n_u + {} 
	\\ \qquad n_\xi + 1.
	\end{split}
	\end{cases}
\end{equation}}

\vspace*{-0.5cm}
We will also use the following alternative and equivalent representation for polynomial NARMAX models:
\begin{equation}
	y_k = \sum_{i=1}^p c_i \prod_{j=0}^{n_u} u_{k-j}^{b_{i,j}} \prod_{l=1}^{n_{\xi}} \xi_{k-l}^{d_{i,l}} \prod_{m=1}^{n_y} y_{k-m}^{a_{i,m}} + \xi_k, \label{eq:NARMAX2}
\end{equation}
where $p$ is the number of model terms, $c_i$ are the model parameters, $a_{i,m}, b_{i,j},d_{i,l} \in \mathbb{N}$ are the exponents for output, input and noise terms.
%
\vspace*{-0.15cm}
\subsection{Proposed TAG representation}
\vspace*{-0.15cm}
\begin{figure} [!tb]
		\centering
			\includegraphics[scale=0.450]{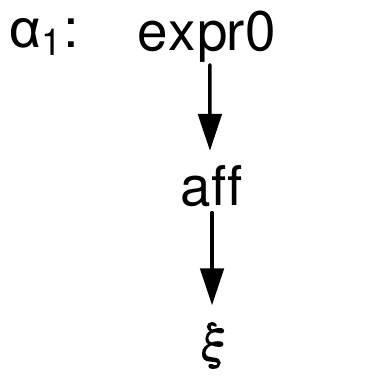}
			\caption{Initial Trees $I$ of TAG $G_\mathrm{N}$.}
			\label{fig:TAG_narmaxI}
\end{figure}
\begin{figure} [!tb]
		\centering
			\includegraphics[scale=0.45]{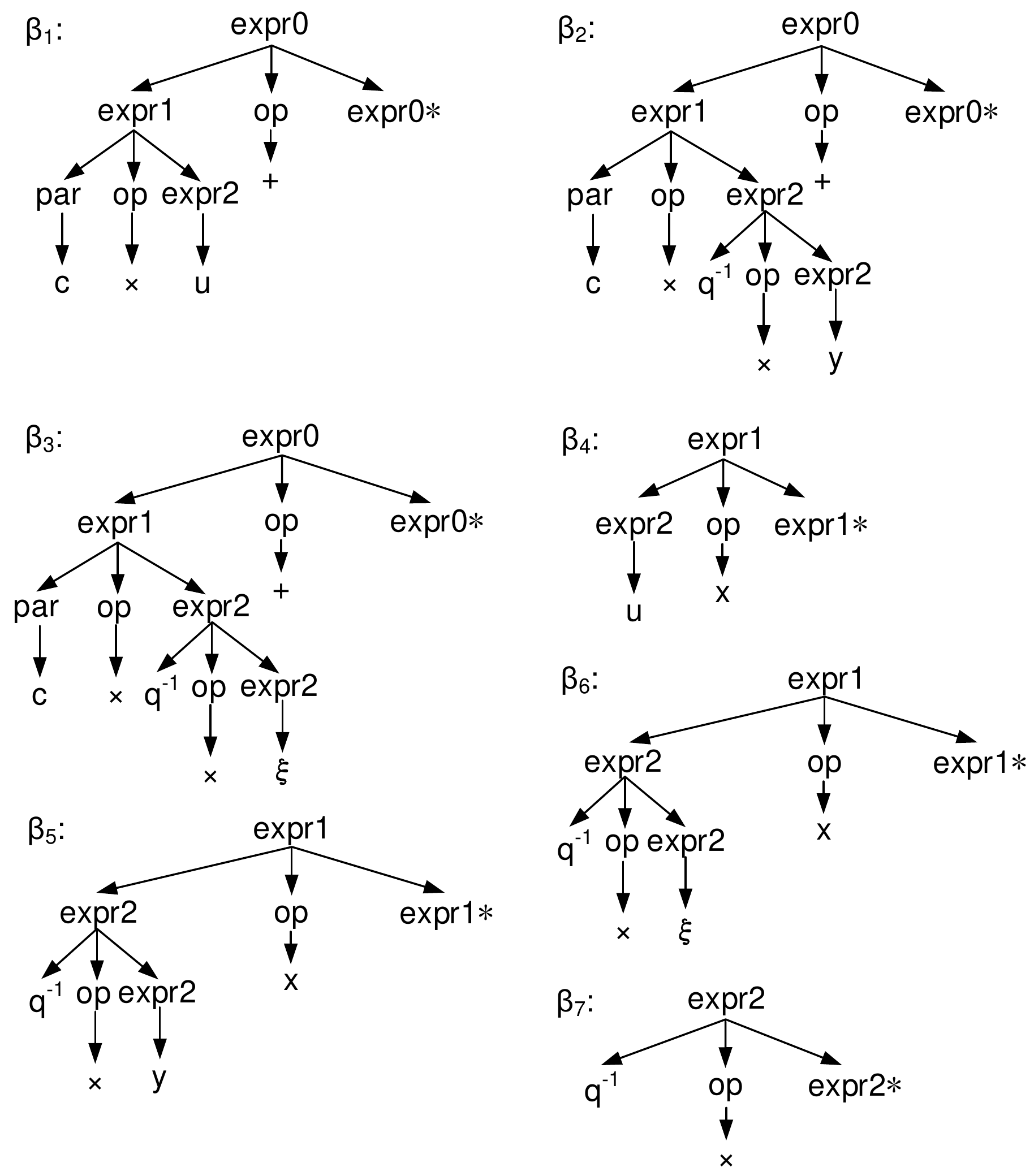}
			\caption{Auxiliary trees $A$ of TAG $G_\mathrm{N}$.}
			\label{fig:TAG_narmaxA}
\end{figure}
In this section we propose a TAG for the polynomial NARMAX model class. The proposed TAG captures the structural relationships in \eqref{eq:NARMAX2}. In the sequel, the time index will be dropped in the context of the proposed TAG, as $q^{-1}$ will be used to denote a backward time shift. For convenience, introduce the following notation. For a given model in the form of \eqref{eq:NARMAX2}, define $J_i \coloneqq \left\{ j \in \mathbb{N}_{\ge 0} \mid b_{i,j} \ne 0 \right\}$ , $L_i \coloneqq \left\{ l \in \mathbb{N}_{> 0} \mid d_{i,l} \ne 0 \right\}$ and $M_i \coloneqq \left\{ m \in \mathbb{N}_{> 0} \mid a_{i,m} \ne 0 \right\}$. For the $i^\text{th}$ model term, the sequence of delays in the input, noise and output factors are denoted by $\left( \bar{j}_n^{(i)} \right)_{n \in J_i}$, $\left( \bar{l}_n^{(i)} \right)_{n \in L_i}$, $\left( \bar{m}_n^{(i)} \right)_{n \in M_i}$ respectively.


\begin{theorem}
	\label{prop:TAG_NARMAX}
	Consider the TAG $G_\mathrm{N} = <N,T,S,I,A>$ with
	\begin{itemize}
		\item[-] $N = \{\mathrm{expr0, expr1, expr2, op, par} \}$,
		\item[-] $T = \{ \mathrm{u, y, \xi, +, c, } \times \mathrm{,} \ q^{-1} \}$,
		\item[-] $S = \mathrm{expr0}$,
		\item[-] $I = \{\alpha_1\}$, where initial tree $\alpha_1$ is depicted in Fig. \ref{fig:TAG_narmaxI},
		\item[-] $A = \{\beta_1, \beta_2, \beta_3, \beta_4, \beta_5, \beta_6, \beta_7 \}$, where the auxiliary trees $\beta_i$'s are depicted in Fig. \ref{fig:TAG_narmaxA}.
	\end{itemize}
	The model set $\mathcal{M}(G_\mathrm{N})$ is equivalent to the set of all models that can be expressed as \eqref{eq:NARMAX2} with finite values of $p,n_u,n_y$ and $n_\xi$.
\end{theorem}
\vspace*{-0.30cm}
\begin{pf}
For the first part of the proof, we show that for any polynomial NARMAX model in the form of \eqref{eq:NARMAX2}, there exists a derivation tree such that the resulting derived tree has a yield that is equal to the RHS of \eqref{eq:NARMAX2}. Algorithm \ref{alg:derivationConstruction} constructs such a derivation tree for a given polynomial NARMAX model. The procedure \textproc{Delays}$(\gamma,v,n)$ adjoins $n$ auxiliary tree $\beta_7$ to the derivation tree $\gamma$ at vertex $v$. The algorithm constructs the derivation tree by introducing the {\color{white}\hlbred{first factor}} ($u,y$ or $\xi$) of each of the $p$ model terms, and subsequently building each of the branches by introducing the remaining {\color{white}\hleme{factors}} with the corresponding {\color{white}\hlmblue{delays}} and {\color{white}\hleme{exponents}}.

\begin{algorithm}[t]
	\caption{Parse NARMAX model \eqref{eq:NARMAX2} to derivation tree.}
	\label{alg:derivationConstruction}
	\begin{algorithmic}[1]
	{\small
		\Require $p, J_i, L_i, M_i, \left( \bar{j}_n^{(i)}\right)_{n \in J_i}, \left( \bar{l}_n^{(i)} \right)_{n \in L_i}$, $\left( \bar{m}_n^{(i)} \right)_{n \in M_i}$
		\State $V \gets \{v_0\}; \ \ l(v_0) \gets \alpha_1$ \Comment[\small]{initialize with start tree}
		\State $r \gets v_0$
		\algrenewcommand{\alglinenumber}[1]{{\setlength{\fboxsep}{1pt}\colorbox{BrickRed}{\color{white}\footnotesize#1:}}}
		\State $V \gets \bigcup_{i=1}^p \{v_{i,1} \} \cup V$ \Comment[\small]{Insert $p$ vertices to begin the $p$ summation branches}
		\State $E \gets \bigcup_{i=2}^p \{ \langle v_{i-1,1}, v_{i,1} \rangle \} \cup \{ \langle v_0, v_{1,1} \rangle \}$
		\algrenewcommand{\alglinenumber}[1]{\color{Black}\footnotesize#1:}
		\For{$i \gets 1,p$}
			\If{$J_i \neq \phi$} \Comment[\small]{If there is an input factor in the $i^\mathrm{th}$ term}
				\algrenewcommand{\alglinenumber}[1]{{\setlength{\fboxsep}{1pt}\colorbox{BrickRed}{\color{white}\footnotesize#1:}}}
				\State $l(v_{i,1}) \gets \beta_1$ \Comment[\small]{For each summation branch, assign the appropriate label to the first vertex}
				\algrenewcommand{\alglinenumber}[1]{{\setlength{\fboxsep}{1pt}\colorbox{MidnightBlue}{\color{white}\footnotesize#1:}}}
				\State $\langle V,E,r \rangle \gets \Call{Delays}{\langle V,E,r \rangle, v_{i,1}, \bar{j}_1^{(i)}}$ \Comment[\small]{Adjoin delay trees}
				\algrenewcommand{\alglinenumber}[1]{\color{Black}\footnotesize#1:}
				\State $b_{i,\bar{j}^{(i)}_1} \gets b_{i,\bar{j}^{(i)}_1} -1$ \Comment[\small]{Reduce the corresponding exponent by 1}
			\ElsIf{$L_i \ne \phi$}
				\algrenewcommand{\alglinenumber}[1]{{\setlength{\fboxsep}{1pt}\colorbox{BrickRed}{\color{white}\footnotesize#1:}}}
				\State $l(v_{i,1}) \gets \beta_3$
				\algrenewcommand{\alglinenumber}[1]{{\setlength{\fboxsep}{1pt}\colorbox{MidnightBlue}{\color{white}\footnotesize#1:}}}
				\State $\langle V,E,r \rangle \gets \Call{Delays}{\langle V,E,r \rangle, v_{i,1}, \bar{l}_1^{(i)}}$
				\algrenewcommand{\alglinenumber}[1]{\color{Black}\footnotesize#1:}
				\State $d_{i,\bar{l}^{(i)}_1} \gets d_{i,\bar{l}^{(i)}_1} -1$
			\ElsIf{$M_i \ne \phi$}
				\algrenewcommand{\alglinenumber}[1]{{\setlength{\fboxsep}{1pt}\colorbox{BrickRed}{\color{white}\footnotesize#1:}}}
				\State $l(v_{i,1}) \gets \beta_2$
				\algrenewcommand{\alglinenumber}[1]{{\setlength{\fboxsep}{1pt}\colorbox{MidnightBlue}{\color{white}\footnotesize#1:}}}
				\State $\langle V,E,r \rangle \gets \Call{Delays}{\langle V,E,r \rangle, v_{i,1}, \bar{m}_1^{(i)}}$
				\algrenewcommand{\alglinenumber}[1]{\color{Black}\footnotesize#1:}
				\State $a_{i,\bar{m}^{(i)}_1} \gets a_{i,\bar{m}^{(i)}_1} -1$
			\EndIf
			\State $s_i \gets 1$ \Comment[\small]{Counter for multiplying remaining factors}
			\ForAll{$j \in J_i$} 
				\algrenewcommand{\alglinenumber}[1]{{\setlength{\fboxsep}{1.0pt}\colorbox{OliveGreen}{\color{white}\footnotesize#1:}}}
				\State $V \gets \bigcup_{n=1}^{b_{i,j}} \{ v_{i,s_i+n,1} \cup V\}; \ \ l( v_{i,s_i+n,1}) \gets \beta_4$
				\State $E \gets \bigcup_{n=1}^{b_{i,j}} \{ \langle v_{i,s_i+n-1,1}, v_{i,s_i+n,1} \rangle  \} \cup E $
				\algrenewcommand{\alglinenumber}[1]{\color{Black}\footnotesize#1:}
				\For{$n \gets 1,\bar{b}_j^{(i)}$} \Comment[\small]{Adjoin delays for multiple factors}
					\algrenewcommand{\alglinenumber}[1]{{\setlength{\fboxsep}{1pt}\colorbox{MidnightBlue}{\color{white}\footnotesize#1:}}}
					\State $ \langle V,E,r \rangle \gets \Call{Delays}{\langle V,E,r \rangle, v_{i,s_i+n,1}, j} $
					\algrenewcommand{\alglinenumber}[1]{\color{Black}\footnotesize#1:}
				\EndFor
				\State $s_i \gets s_i + \bar{b}_j^{(i)}$
			\EndFor
			\ForAll{$l \in L_i$}
				\algrenewcommand{\alglinenumber}[1]{{\setlength{\fboxsep}{1.0pt}\colorbox{OliveGreen}{\color{white}\footnotesize#1:}}}
				\State $V \gets \bigcup_{n=1}^{d_{i,l}} \{ v_{i,s_i+n,1} \cup V\}; \ \ l( v_{i,s_i+n,1}) \gets \beta_6$
				
				\State $E \gets \bigcup_{n=1}^{d_{i,l}} \{ \langle v_{i,s_i+n-1,1}, v_{i,s_i+n,1} \rangle  \} \cup E $
				\algrenewcommand{\alglinenumber}[1]{\color{Black}\footnotesize#1:}
				\For{$n \gets 1,\bar{d}_l^{(i)}$}
					\algrenewcommand{\alglinenumber}[1]{{\setlength{\fboxsep}{1pt}\colorbox{MidnightBlue}{\color{white}\footnotesize#1:}}}
					\State $ \langle V,E,r \rangle \gets \Call{Delays}{\langle V,E,r \rangle, v_{i,s_i+n,1}, l} $
					\algrenewcommand{\alglinenumber}[1]{\color{Black}\footnotesize#1:}
				\EndFor
				\State $s_i \gets s_i + \bar{d}_l^{(i)}$
			\EndFor
			\ForAll{$m \in M_i$}
				\algrenewcommand{\alglinenumber}[1]{{\setlength{\fboxsep}{1.0pt}\colorbox{OliveGreen}{\color{white}\footnotesize#1:}}}
				\State $V \gets \bigcup_{n=1}^{a_{i,m}} \{ v_{i,s_i+n,1} \cup V\}; \ \ l( v_{i,s_i+n,1}) \gets \beta_5$
				\State $E \gets \bigcup_{n=1}^{a_{i,m}} \{ \langle v_{i,s_i+n-1,1}, v_{i,s_i+n,1} \rangle  \} \cup E $
				\algrenewcommand{\alglinenumber}[1]{\color{Black}\footnotesize#1:}
				\For{$n \gets 1,\bar{a}_m^{(i)}$}
					\algrenewcommand{\alglinenumber}[1]{{\setlength{\fboxsep}{1pt}\colorbox{MidnightBlue}{\color{white}\footnotesize#1:}}}
					\State $\langle V,E,r \rangle \gets \Call{Delays}{\langle V,E,r \rangle, v_{i,s_i+n,1}, m}$
					\algrenewcommand{\alglinenumber}[1]{\color{Black}\footnotesize#1:}
				\EndFor
				\State $s_i \gets s_i + \bar{a}_m^{(i)}$
			\EndFor
		\EndFor
		\Return $\langle V,E,r \rangle$
		}
	\end{algorithmic}
\end{algorithm}

For the second part of the proof, it needs to be shown that all expressions in $L(G_\mathrm{N})$, i.e., yields of all possible trees generated by $G_\mathrm{N}$, are RHS expressions of polynomial NARMAX models. This is proven by structural induction. We first observe that the simplest tree in $L(G_\mathrm{N})$ is the initial tree $\alpha_1$ with the yield $\xi$. This corresponds to the model
\begin{equation}
	y_k = \xi_k,
\end{equation}
which belongs to the polynomial NARMAX class. Now, consider an arbitrary saturated derived tree $\gamma \in L_\mathrm{T}(G_\mathrm{N})$ whose yield is the RHS of a polynomial NARMAX model. This implies that the yield is a polynomial expression in terms of the factors $u$, $y$ and $\xi$. To complete the principle of induction, it must be shown that any possible adjunction to $\gamma$ results in a new tree in $L_\mathrm{T}(G_\mathrm{N})$ whose yield is also a polynomial expression in terms of the aforementioned factors.

For convenience, the auxiliary trees are grouped based on the operators involved - $\beta_1, \beta_2, \beta_3$ are called \textit{additive-type} auxiliary trees, $\beta_4, \beta_5, \beta_6$ are called \textit{multiplicative-type}, and $\beta_7$ is called \textit{delay-type} auxiliary tree. The following adjunctions be made on $\gamma$:
\begin{itemize}
	\item adjunction of an additive-type tree. Such an adjuction introduces an input, output or noise term additively in the expression while respecting the causality of the expression. Hence the resulting expression is also a polynomial;
	\item adjunction of a multiplicative-type tree. This simply introduces multiplicative factors to an existing model term, and hence, the resulting expression is also a polynomial;
	\item adjunction of a delay-type tree. This operation simply adds delays to an existing monomial, and hence preserves the polynomial structure of the expression.
\end{itemize}
Since all possible operations yield a causal polynomial expression, it can be concluded that $L(G_\mathrm{N})$ consists of only dynamical polynomial expressions in terms of the factors $u,y$ and $\xi$ which corresponds to a polynomial NARMAX model. This concludes the proof. \hfill \qed
\end{pf}
\vspace*{-0.30cm}

Theorem \ref{prop:TAG_NARMAX} demonstrates that structural properties of a rich class of dynamical models can be captured within a compact set of trees of a TAG. The expansive representational capability of TAG can be exploited using EAs such as GP to identify models without prior specification of structure and complexity, as demonstrated in \cite{khandelwal2019grammar}. Furthermore, Algorithm \ref{alg:derivationConstruction} provides a method to compute the derivation tree representation of a given polynomial NARMAX model in terms of grammar $G_\mathrm{N}$. Consequently, available prior information about the model of the system can be translated to TAG representation (or incorporated in tree sets $I, A$), thereby making the evolutionary search more efficient. Hence, the use of TAG enables identification within a larger class of dynamical models without requiring user-interaction, while simultaneously allowing the user to restrict the evolutionary search effectively.

\begin{figure*} [!tb]
		\centering
		\begin{subfigure}[t]{0.45\linewidth}
			\centering
			\includegraphics[scale=0.55]{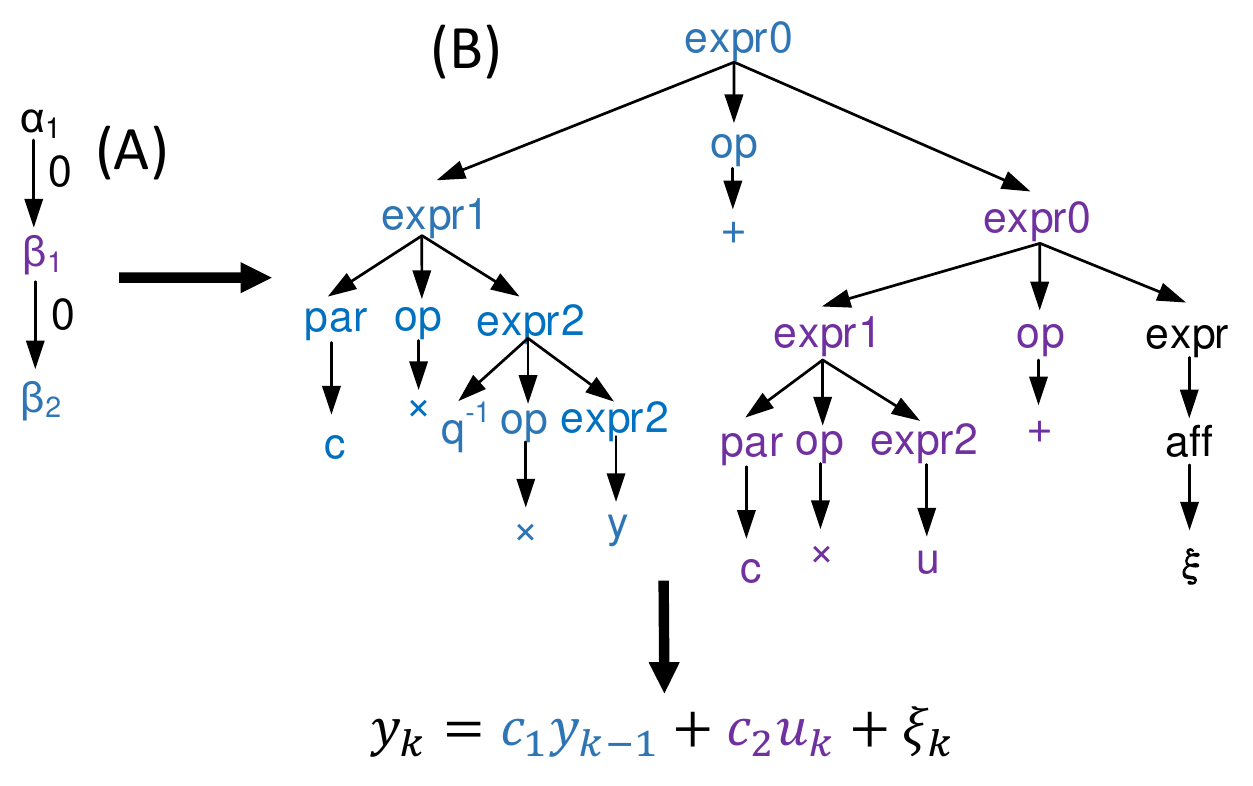}
			\caption{Example 1 - derivation tree (A), derived tree (B) and symbolic model.}
			\label{fig:TAG_ARX_ex}
		\end{subfigure}
		~%
		\begin{subfigure}[t]{0.45\linewidth}
			\centering
			\includegraphics[scale=0.55]{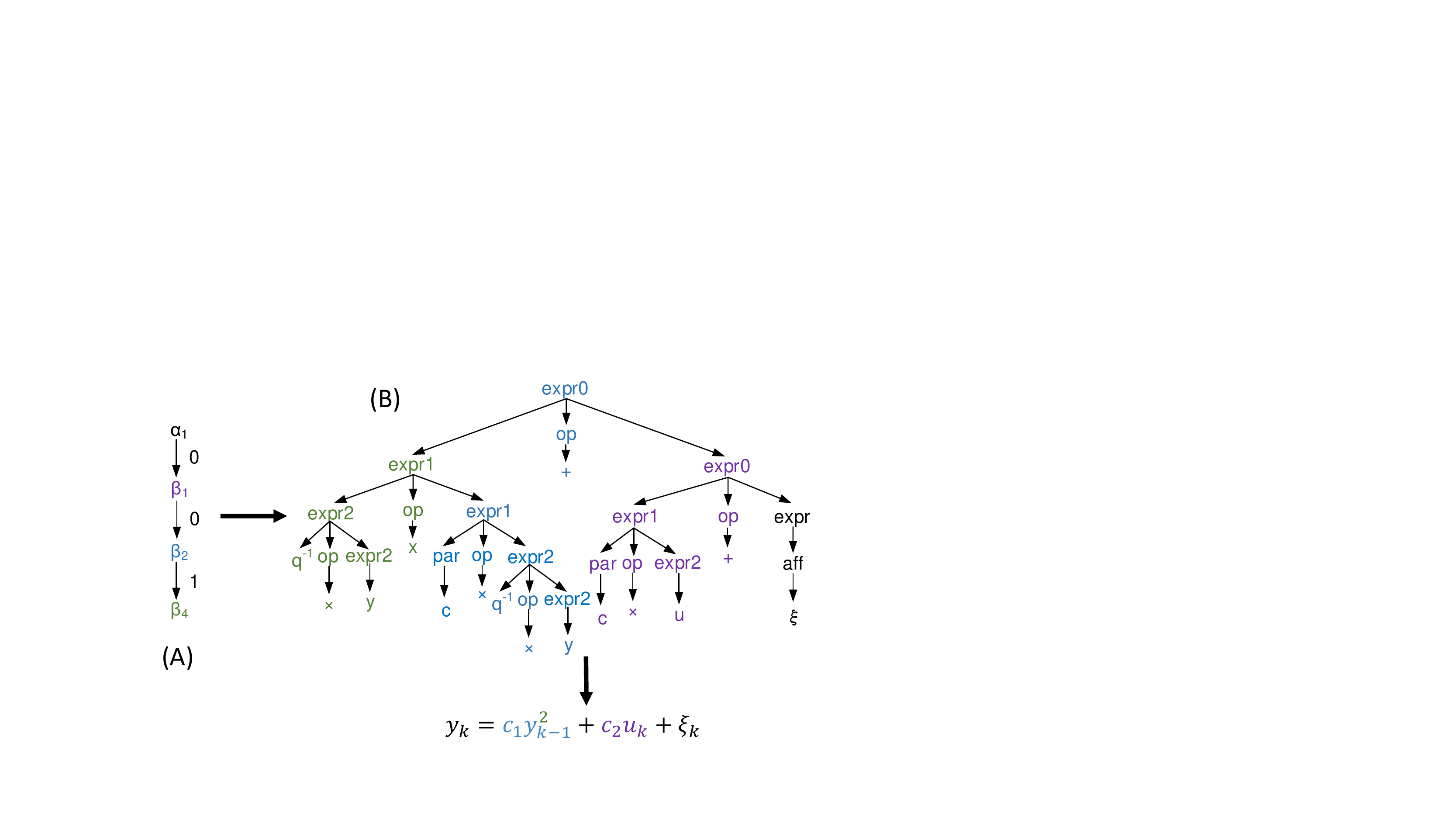}
			\caption{Example 2 - derivation tree (A), derived tree (B) and symbolic model.}
			\label{fig:TAG_NARX_ex}
		\end{subfigure}
		\\%
		\begin{subfigure}[t]{\linewidth}
			\centering
			\includegraphics[scale=0.60]{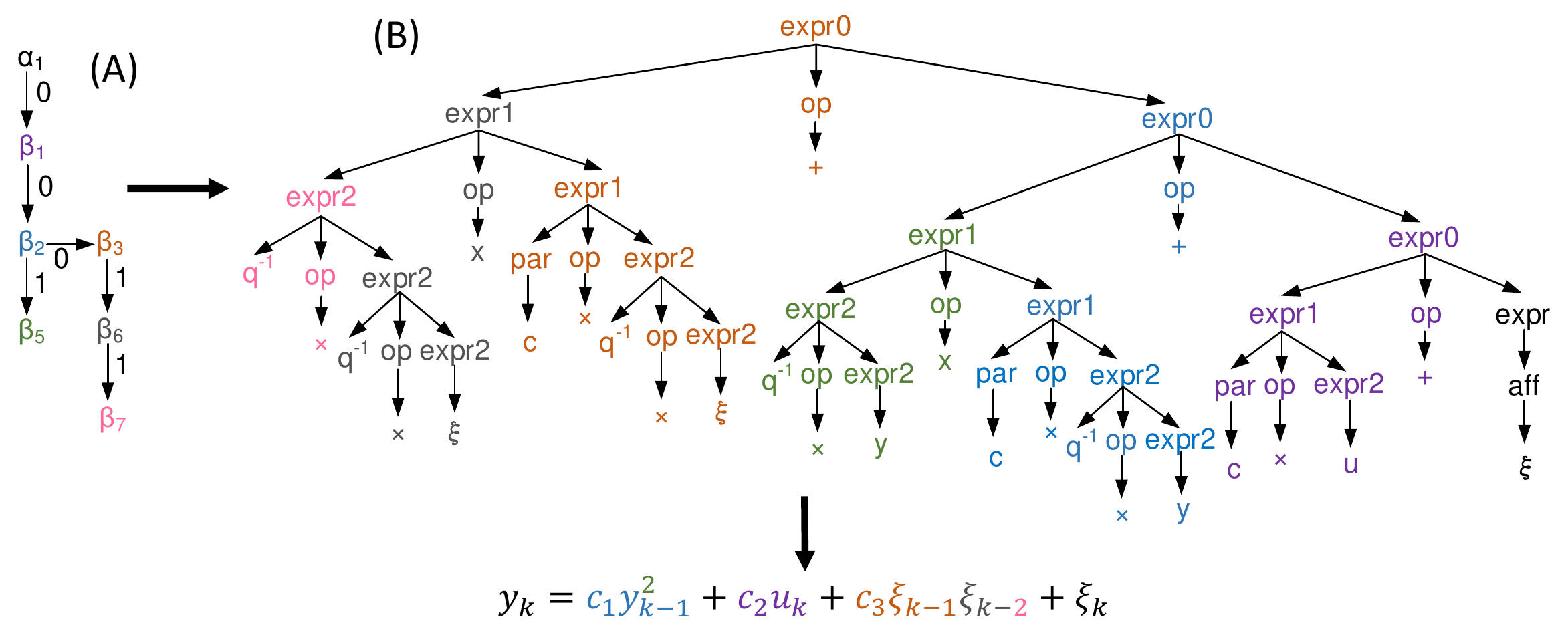}
			\caption{Example 3 - derivation tree (A), derived tree (B) and symbolic model.}
			\label{fig:TAG_NARMAX_ex}
		\end{subfigure}
		\caption{Illustrative examples.}
		\label{fig:illustrations}
\end{figure*}

\vspace*{-0.15cm}
\section{Illustrations}
\label{sec:illustrations}
\vspace*{-0.1cm}
In this section we discuss aspects of TAG useful for EA-based SI. We demonstrate the use of TAG $G_\mathrm{N}$ to generate polynomial NARMAX models. It is also shown that models belonging to simpler model classes can be generated by scaling down the set of elementary trees of $G_\mathrm{N}$ appropriately. Furthermore, more flexible model classes can be represented by scaling up the set of elementary trees. This is demonstrated by extending the proposed TAG to generate Non-linear Box Jenkins (NBJ) models.

\vspace*{-0.15cm}
\subsection{Model generation using $G_\mathrm{N}$}
\vspace*{-0.15cm}
Three illustrative examples are used to demonstrate the generation of models using $G_\mathrm{N}$. The models generated belong to the ARX, polynomial NARX and polynomial NARMAX model classes. It will be demonstrated that by restricting the elementary trees $I$ and $A$ to subsets of the elementary trees in the proposed TAG $G_\mathrm{N}$, we can generate models that only belong to model sub-classes that are properly included in the set of polynomial NARMAX models, such as FIR and truncated Volterra series.

\vspace*{-0.10cm}
\subsubsection{ARX example}
\vspace*{-0.10cm}
ARX models can be described by the equation
\vspace*{-0.10cm}
\begin{equation}
	y_k = \sum_{i=0}^{n_u} b_i u_{k-i} + \sum_{j=1}^{n_y} a_j y_{k-j} + \xi_k,
\end{equation}
where $a_j, b_i \in \mathbb{R}$ are coefficients. The grammar $G_\mathrm{N}$ can be used to generate ARX models by restricting the auxiliary tree set $A$ as
\vspace*{-0.10cm}
\begin{equation}
	A'=\{ \beta_1, \beta_2, \beta_7 \} \subset A.
\end{equation}
Consider the example depicted in Fig. \ref{fig:TAG_ARX_ex}. Tree (A) is a derivation tree with initial tree $\alpha_1$ at the root node, and auxiliary trees $\beta_1$ and $\beta_2$ in subsequent vertices. The edges are labelled with Gorn addresses of vertices in the auxilliary trees at which adjunctions take place. Performing the adjunctions results in derived tree (B) in Fig. \ref{fig:TAG_ARX_ex}. The RHS of the resulting model appears at the leaves of the derived tree, and the corresponding model is 
\vspace*{-0.10cm}
\begin{equation}
	y_k = c_1 y_{k-1} + c_2 u_k + \xi_k.
\end{equation}
\vspace*{-0.10cm}
\vspace*{-0.30cm}
\subsubsection{NARX example}
\vspace*{-0.10cm}
Polynomial NARX models can be described by the equation
\begin{equation}
	y_k = \sum_{i=1}^p c_i \prod_{j=0}^{n_u} u_{k-j}^{b_{i,j}} \prod_{m=1}^{n_y} y_{k-m}^{a_{i,m}} + \xi_k.
\end{equation}
By restricting auxiliary trees to the set
\vspace*{-0.05cm}
\begin{equation}
	A''=\{ \beta_1, \beta_2, \beta_4, \beta_5, \beta_7 \} \subset A
\end{equation}
we can restrict the proposed grammar to generate polynomial NARX models only. Consider the example derivation tree (A) in Fig. \ref{fig:TAG_NARX_ex}, which is an extension of the previous example. The derivation tree consists of the initial tree $\alpha_1$, and auxiliary trees $\beta_2, \beta_3$ and $\beta_4$. Performing the adjunctions described by the derivation tree results in the derived tree (B) in Fig. \ref{fig:TAG_NARX_ex}. The corresponding symbolic model is
\vspace*{-0.10cm}
\begin{equation}
	y_k = c_1 y_{k-1}^2 + c_2 u_k + \xi_k.
\end{equation}

\subsubsection{NARMAX example}
\vspace*{-0.10cm}
This example builds on the previous example by using the complete auxiliary tree set $A$ and adjoining trees $\beta_3, \beta_6$ and $\beta_7$ to the tree $\beta_2$. The new derivation tree and derived tree are depicted in Fig. \ref{fig:TAG_NARMAX_ex}. The corresponding model,
\vspace*{-0.10cm}
\begin{equation}
	y_k = c_1 y_{k-1}^2 + c_2 u_k + c_3 \xi_{k-1} \xi_{k-2} \xi_k + \xi_k,
\end{equation}
is a polynomial NARMAX model.

\begin{figure} [!tb]
		\centering
			\includegraphics[scale=0.40]{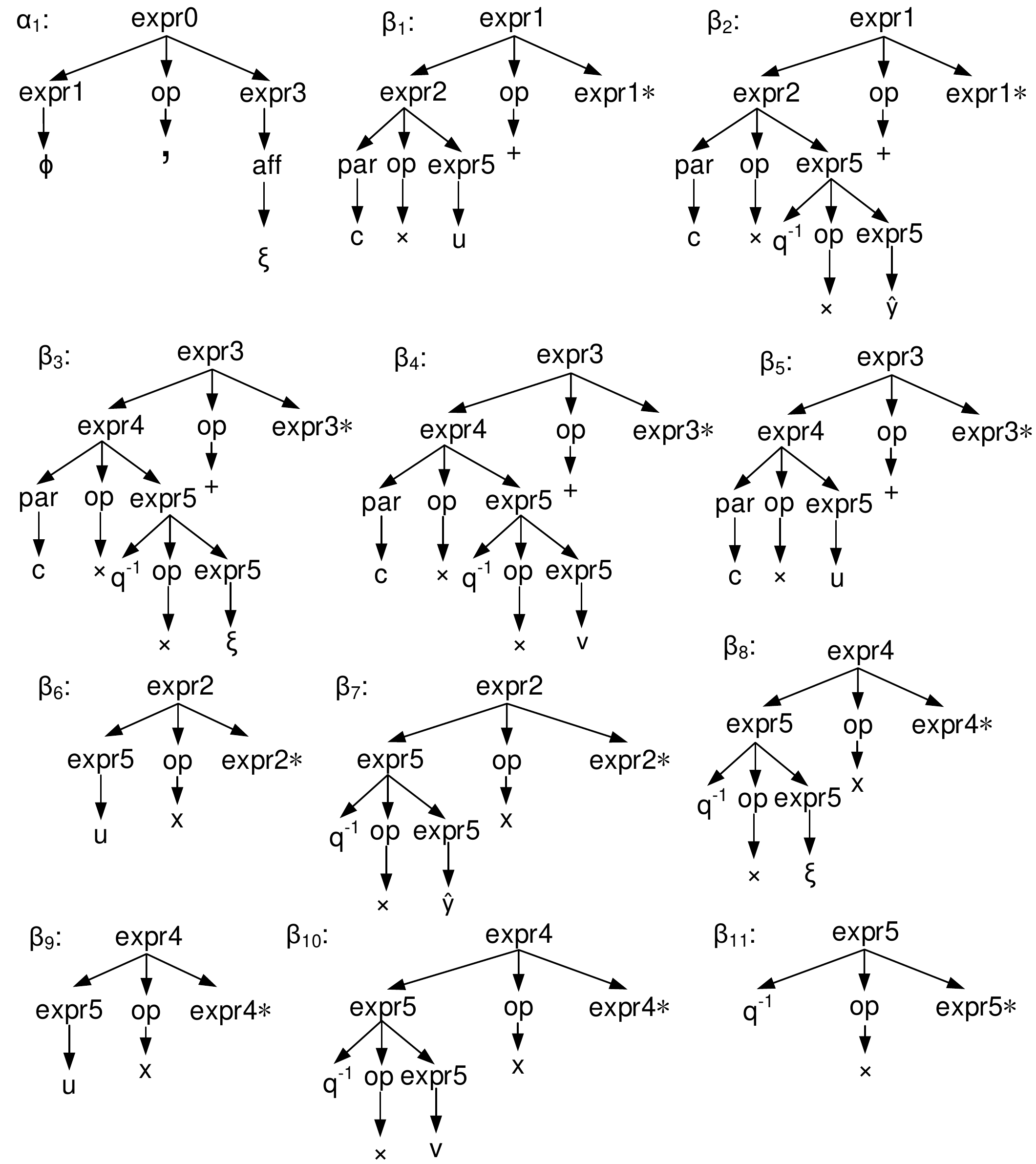}
			\caption{Initial and auxiliary trees ($I$ and $A$) of TAG $G_\mathrm{NBJ}$}
			\label{fig:TAG_NBJ}
\end{figure}

\vspace*{-0.15cm}
\subsection{Non-linear Box-Jenkins Extension}
\vspace*{-0.15cm}
Just like the proposed grammar can be scaled down to generate specific dynamic sub-classes, it can also be extended to generate models that belong to a more generalized class of models. We illustrate this by extending the proposed grammar to a more generalized models structure - Non-linear Box Jenkins (NBJ).

In the case of linear systems, a Box-Jenkins model structure is an extension of the Output Error (OE) model structure, where the error is modelled as an ARMA process \citep{Ljung1999}. The BJ class also includes, as special cases, other linear model structures such as ARMAX and OE. In the same spirit, NBJ model structure can be expressed as a Non-linear Output Error (NOE) model where the error is subsequently modelled as a NARMA process. The NBJ model structure is given by the following equations
\vspace*{-0.7cm}

{\small
\begin{align}
\hat{y}_k &= f(\hat{y}_{k-1}, \dots, \hat{y}_{k-n_y}, u_k, \dots, u_{k-n_u}), \nonumber \\
v_k &= g(v_{k-1}, \dots, v_{k-n_v}, u_{k}, ..., u_{k-n_u}, \xi_{k-1}, \dots, \xi_{k-n_{\xi}}) + \xi_k, \nonumber \\
y_k &= \hat{y}_k + v_k,
\end{align}
}

\vspace*{-0.7cm}
\noindent
where $f(\cdot)$ and $g(\cdot)$ are polynomial functions in terms of their arguments. Notice that the RHS expressions of the equations describing the process and noise dynamics have the same structure that was studied in Sec. \ref{sec:NARMAX_TAG} for NARMAX models (see \eqref{eq:NARMAX2}). Hence, the proposed TAG can be extended to generate NBJ models. Fig. \ref{fig:TAG_NBJ} depicts the initial and auxiliary trees of the grammar for NBJ model structures $G_\mathrm{NBJ}$. The structure of the initial tree $\alpha_1$ ensures that all elements in $L(G_\mathrm{NBJ})$ contain two expressions, separated by a comma, that represent the functions $f(\cdot)$ and $g(\cdot)$ respectively. Each of these expressions can be expanded by adjoining auxiliary trees that ensure that the polynomial structure is maintained.

\vspace*{-0.15cm}
\section{Conclusions}
\label{sec:conclusions}
\vspace*{-0.1cm}

We presented a TAG based concept of a model set, that is more general than that commonly used in the system identification literature. A TAG $G_\mathrm{N}$ was proposed that captures the dynamical structure of polynomial NARMAX models. It was demonstrated that sub-classes of the polynomial NARMAX class can be represented by choosing an appropriate subset of the elementary trees of $G_\mathrm{N}$. Similarly, more flexible model classes like Non-linear Box-Jenkins can be represented by extending the set of elementary trees. This illustrates that a compact set of elementary trees can be used to express the dynamical relationships across a variety of model classes, thereby enabling the design of TAG-based EA approaches for SI that require minimal user-interaction. The practical soundness of this concept has been demonstrated in \cite{khandelwal2019grammar}, where a TAG-based EA approach was used to identify a non-linear benchmark dataset with minimal user-interaction, and also in \cite{khandelwal2019data}, where the same TAG-based EA approach is used to identify multiple real physical systems and benchmark data set with minimal changes in the methodology itself.

\bibliographystyle{apa}
\vspace*{-0.2cm}
\bibliography{grammar01}

\end{document}